*To be submitted to Phys. Rev. B*

# Thermodynamics of Electromechanically-Coupled Mixed Ionic-Electronic Conductors: Deformation potential, Vegard strains and Flexoelectric effect

A.N. Morozovska,[1,*] E.A. Eliseev,[1,2], A.K. Tagantsev[3], S.L. Bravina[4],
Long-Qing Chen[5] and S.V. Kalinin[6,†]

[1] Institute of Semiconductor Physics, National Academy of Science of Ukraine,
41, pr. Nauki, 03028 Kiev, Ukraine

[2] Institute for Problems of Materials Science, National Academy of Science of Ukraine,
3, Krjijanovskogo, 03142 Kiev, Ukraine

[3]Ceramics Laboratory, Swiss Federal Institute of Technology (EPFL),
CH-1015 Lausanne, Switzerland

[4] Institute of Physics, National Academy of Science of Ukraine,
46, pr. Nauki, 03028 Kiev, Ukraine

[5]Department of Materials Science and Engineering, Pennsylvania State University,
University Park, Pennsylvania 16802, USA

[6] The Center for Nanophase Materials Sciences and Materials Sciences and Technology Division, Oak Ridge National Laboratory, Oak Ridge, TN 37831

[*]morozo@i.com.ua

[†] sergei2@ornl.gov

Strong coupling among external voltage, electrochemical potentials, concentrations of electronic and ionic species, and strains is a ubiquitous feature of solid state mixed ionic-electronic conductors (MIECs), the materials of choice in devices ranging from electroresistive and memristive elements to ion batteries and fuel cells. Here, we analyze in detail the electromechanical coupling mechanisms and derive generalized bias-concentration-strain equations for MIECs including effects of concentration-driven chemical expansion, deformation potential, and flexoelectric effect contributions. This analysis is extended towards the bias-induced strains in the uniform and scanning probe microscopy-like geometries. Notably, the contribution of the electron-phonon and flexoelectric coupling to the local surface displacement of the mixed ionic-electronic conductor caused by the electric field scanning probe microscope tip has not been considered previously. The developed thermodynamic approach allows evolving theoretical description of mechanical phenomena induced by the electric fields (electro-mechanical response) in solid state ionics towards analytical theory and phase-field modeling of the MIECs in different geometries and under varying electrical, chemical, and mechanical boundary conditions.

# 1. Introduction

Development of strains is a phenomenon ubiquitous in solid-state electrochemical devices including batteries [1, 2], fuel cells [3, 4] and electroresistive and memristive electronics. For example, strain is one of the dominant factors contributing to the mechanical instability of solid oxide fuel cells and Li-ion battery anodes such as intra-particle cracking and delamination of electrodes [5, 6]. The difference in boundary conditions (clamped or unclamped material) can significantly shift the electrochemical potentials of reacting species and electrons [7] and affect charge-discharge hysteresis and hence efficiency of materials and devices. On the other hand, electrochemically generated strains can be utilized to build electromechanical devices such as artificial muscles [8] and actuators [9], or diagnostic tool for electrochemical systems at both the macroscopic [10] and nanometer scales [11]. Electrochemical Strain Microscopy [11, 12] uses the periodic nanoscale electrochemical strains generated by a biased scanning probe of microscope to detect Li-ion diffusion in cathode [13] and anode materials [14] at the 10-100 nanometer scale. Based on the previous imaging and spectroscopy results in ferroelectric materials [15, 16, 17] it is possible to perform electrochemical strain microscopy measurements at the level of several nanometers, opening the pathway for probing structure-electrochemical property relationships at a single structural defect.

A common source for strain in electrochemically active materials is the compositional dependence of lattice parameters, as discussed in detail by Larche and Cahn [18]. This is the case for many ionic and mixed ionic-electronic conductors such as ceria [19], cobaltites [20, 21, 22, 23], nikelates [24] and manganites [25]. Similarly, insertion and extraction of Li-ions in Li-battery electrodes produce large volume changes [26, 27]. Most of the previous theoretical studies of strain effects in diffusional [28, 29] and electrochemical systems consider this compositional lattice expansion as the only source of strain. This assumption is reasonable if the electronic conductivity of a material is sufficiently high to avoid significant potential drops (equivalent to the presence of support electrolyte in liquid electrochemistry [30, 31]), obviating electromigration transport and providing local electroneutrality.

However, the situation can differ significantly for the case of materials with finite electronic conductivity, in which both concentration fields and electrostatic field are non-uniform within the material. Electrostatic fields in the material give rise to strains due to electrostriction [32, 33, 34, 35] and space-charge [36] effects. Secondly, the changes in the redox state of Jahn-Teller (JT) active cations can give rise to additional strain coupling mechanisms through the deformation potential [37, 38, 39, 40, 41, 42]. As an example, in perovskites these effects can be understood as a consequence of the changes in favored oxygen octahedral geometry as a function of oxidation state of the central cation. Similarly to the fact that change in the *d*-orbital population changes octahedral shape and gives rise to JT effect, the strain deforming octahedral will shift the electrochemical potential of the central atom. These effects will be particularly pronounced on the nanometer scale as relevant to scanning probe microscopy imaging [43] and nanoparticle/nanowire materials, in which the conditions of local electroneutrality are violated on the length scales of corresponding screening lengths and large (compared to macroscopic systems) strains can be supported.

Inhomogeneous electric fields, which are inevitably present in systems with inhomogeneous space charge (e.g. in the vicinity of the tip-surface junction), induces elastic strains linearly proportional to the field gradient due to the flexoelectric coupling; vice versa inhomogeneous elastic stress causes electric polarization. The existence of such effect was pointed out by Mashkevich and Tolpygo [44] and Kogan [45]. A comprehensive theory of the flexoelectric effect was offered by Tagantsev [46, 47, 48], experimental measurements of flexoelectric tensor components in bulk crystals were for perovskites carried out by Ma and Cross [49, 50, 51, 52, 53] and Zubko et al. [54]. Further theoretical developments of the flexoelectric response of different nanostructures were made by Catalan et al [55, 56], Majdoub et al. [57], Kalinin and Meunier [58], Eliseev et al [59], and Sharma et al [60, 61].

In this paper, we develop the equilibrium strain-concentration-bias equations for electrochemically active materials that account both for chemical expansivity, deformation potential and flexoelectric effects. The relevant comparison here is the Ginzburg-Landau type theories for ferroelectric materials that are broadly available for ferroelectrics and allow domain structures [63], domain dynamics [62], behavior in non-uniform systems (e.g. strained films and multilayers [63]) and the effects of individual and multiple defects to be explored [64]. Once available for electrochemical systems, similar advances based on phase-field type models could be achieved [65], [66], [67, 68].

## 2. Generalized concentration-strain-bias constitutive relation

Here, we analyze the coupling between electrochemical potential and strain in mixed ionic-electronic conductors (MIEC). We consider the flexoelectric effect, deformation potential, quasi-Fermi levels shift by electron-phonon coupling and Vegard expansion of the lattice caused by mobile donor (and/or acceptors) as the primary contributing mechanisms.

### *2.1. Flexoelectric effect contribution into electrostatic potential and elastic stress*

For centrosymmetric crystals (considered hereinafter) the ***direct*** flexoelectric effect gives the equation of state for dielectric polarization $P_i(\mathbf{r})$ [46, 47]:

$$P_i = \gamma_{klij}\frac{\partial u_{kl}}{\partial x_j} + \varepsilon_0\chi_{ij}E_j, \tag{1}$$

which includes the "flexoelectric" polarization $\gamma_{ijkl}\,\partial u_{jk}/\partial x_l$ induced by inhomogeneous strain $u_{ij}(\mathbf{r})$ gradient, $\partial u_{ij}/\partial x_l$ [47, 53, 54], and dielectric response $\varepsilon_0(\varepsilon_{ij}-\delta_{ij})E_j$, where $\varepsilon_0$ is universal vacuum dielectric constant, $\chi_{ij}=(\varepsilon_{ij}-\delta_{ij})$ is the lattice susceptibility tensor, $\varepsilon_{ij}$ is the lattice permittivity tensor. $E_i$ is the electric field. The flexoelectric strain tensor $\gamma_{ijkl}$ has been measured experimentally for several substances and it was found to vary by several orders of magnitude from $10^{-11}$C/m to $10^{-6}$C/m [69].

Direct substitution of the polarization (1) into Maxwell equation $\mathrm{div}(\mathbf{P}+\varepsilon_0\mathbf{E})=\rho_f$ along with definition $E_k(\mathbf{r})=-\partial\varphi(\mathbf{r})/\partial x_k$ leads to the Poisson-type equation with flexoelectric term for the electric potential $\varphi(\mathbf{r})$ of MIEC:

$$\varepsilon_0\varepsilon_{ij}\frac{\partial^2\varphi(\mathbf{r})}{\partial x_i\partial x_j} = -q\left(p(\mathbf{r}) - n_C(\mathbf{r}) - N_a^-(\mathbf{r}) + N_d^+(\mathbf{r})\right) + \gamma_{ijkl}\frac{\partial^2 u_{ij}(\mathbf{r})}{\partial x_k\partial x_l} \tag{2}$$

Here $q$ is the absolute value of electron charge, $n_C(\mathbf{r})$ is the concentration of electrons in the conduction band, $p(\mathbf{r})$ is the concentration of holes in the valence band, $N_d^+(\mathbf{r})$ is the concentration of mobile ionized donors, and $N_a^-(\mathbf{r})$ is the concentration of mobile ionized acceptors in the MIEC.

The ***converse*** flexoelectric effect contributes into the Hook's law relating the strain $u_{kl}(\mathbf{r})$ and stress tensor $\sigma_{kl}(\mathbf{r})$ [70]:

$$\sigma_{ij}(\mathbf{r}) = c_{ijkl} u_{kl}(\mathbf{r}) + f_{ijkl} \frac{\partial P_k(\mathbf{r})}{\partial x_l}. \tag{3a}$$

here $c_{ijkl}$ is the tensor of elastic stiffness, flexoelectric stress tensor $f_{ijkl} = \gamma_{ijmk} \chi_{ml}^{-1} / \varepsilon_0$. Hereinafter we neglect the contribution of quadratic contribution of the flexoelectric effect and using Eq.(1) rewrite Eq. (3a) as [71]:

$$\sigma_{ij}(\mathbf{r}) = c_{ijkl} u_{kl}(\mathbf{r}) + \gamma_{ijmk} \frac{\partial E_k(\mathbf{r})}{\partial x_m}. \tag{3b}$$

The substitution of the polarization from Eq.(1) into Eq.(3b) leads to the relations:

$$\sigma_{ij}(\mathbf{r}) = c_{ijkl} u_{kl}(\mathbf{r}) - \gamma_{ijkl} \frac{\partial^2 \varphi(\mathbf{r})}{\partial x_k \partial x_l}, \tag{4a}$$

$$u_{ij}(\mathbf{r}) = s_{ijkl} \sigma_{kl}(\mathbf{r}) + s_{ijmn} \gamma_{mnkl} \frac{\partial^2 \varphi(\mathbf{r})}{\partial x_k \partial x_l}. \tag{4b}$$

Where $\gamma_{ijkl} \frac{\partial^2 \varphi(\mathbf{r})}{\partial x_k \partial x_l}$ is the linear contribution of the flexoelectric effect, $s_{ijkl}$ is the tensor of elastic compliances.

### *2.2. Vegard expansion of the lattice caused by mobile donor and acceptors*

Effect of the stoichiometry on the local strain is the *linear* dependence of lattice constants on the chemical composition of solid solution (Vegard law of chemical expansion [18, 72]). In accordance with the Vegard law the local stress $\sigma_{ij}$ and strains $u_{ij}$ produced by the mobile ions (donors or acceptors) migration and diffusion are related as [1, 29]:

$$\sigma_{ij} = c_{ijkl} u_{kl}(\mathbf{r}) - \beta_{ij}^{d} \left( N_d^+(\mathbf{r}) - N_{d0}^+ \right) - \beta_{ij}^{a} \left( N_a^-(\mathbf{r}) - N_{a0}^- \right), \tag{5a}$$

$$u_{ij} = s_{ijkl} \sigma_{kl}(\mathbf{r}) + \tilde{\beta}_{ij}^{d} \left( N_d^+(\mathbf{r}) - N_{d0}^+ \right) + \tilde{\beta}_{ij}^{a} \left( N_a^-(\mathbf{r}) - N_{a0}^- \right), \tag{5b}$$

where $N_d^+(\mathbf{r})$ is the instant concentration of mobile ionized donors, $N_a^-(\mathbf{r})$ is the instant concentration of mobile ionized acceptors, $N_{d0}^+$ and $N_{a0}^-$ are their stoichiometric equilibrium concentrations, $\beta_{ij}^{a,d}$ and $\tilde{\beta}_{ij}^{a,d} = s_{ijkl} \beta_{kl}^{a,d}$ are the *Vegard expansion* tensors for acceptors (donors).

The structure of Vegard expansion tensor is controlled by the symmetry (crystalline or Curie group symmetry) of the material; for isotropic or cubic media it is diagonal and reduces to scalar: $\beta_{jk}^{a,d} = \beta^{a,d}\delta_{jk}$ (hereinafter $\delta_{jk}$ is the Kroneker-delta symbol). Experimental methods for $\beta_{ij}$ determination are relatively well established. For instance, one could either directly study the strain of a given sample with the changes of stoichiometry (see e.g. [23, 24, 25]) or consider the set of several samples with slightly different composition (solid solution).

Note, that the Vegard strain caused by mobile donors and acceptors leads to the shift of their chemical potential levels proportional to the convolution $\beta_{jk}^{a} u_{jk}(\mathbf{r})$ or $\tilde{\beta}_{jk}^{a}\sigma_{jk}(\mathbf{r})$ (see e.g. Ref.[1]) and their equilibrium concentrations in the Boltzmann-Planck-Nernst approximation:

$$N_d^+(\mathbf{r}) \approx N_{d0}^+ \exp\left(\frac{\beta_{jk}^d u_{jk}(\mathbf{r}) - q\varphi(\mathbf{r})}{k_B T}\right), \tag{6a}$$

$$N_a^-(\mathbf{r}) \approx N_{a0}^- \exp\left(\frac{\beta_{jk}^a u_{jk}(\mathbf{r}) + q\varphi(\mathbf{r})}{k_B T}\right). \tag{6b}$$

Where $k_B$=1.3807×10$^{-23}$ J/K, $T$ is the absolute temperature.

Consequently, the Eqs. (5) and (6) can be interpreted as the direct and converse Vegard effect: ions concentration variation induces stress/strain (the ***direct Vegard effect***), where the strain/stress produces the concentration changes (the ***converse Vegard effect***).

### *2.3. Electron-phonon coupling contribution in elastic subsystem*

In deformation potential theory [37-42], the strain induced conduction (valence) band edge shift is proportional to the strain in the linear approximation, namely:

$$E_C\left(u_{ij}(\mathbf{r})\right) = E_C(0) + \Xi_{ij}^C u_{ij}(\mathbf{r}), \qquad E_V\left(u_{ij}(\mathbf{r})\right) = E_V(0) - \Xi_{ij}^V u_{ij}(\mathbf{r}). \tag{7}$$

where $E_C$ and $E_V$ are the energetic position of the bottom of conduction band and the top of the valence band respectively [73], $\Xi_{ij}^{C,V}$ is a tensor deformation potential of electrons in the conduction ($C$) and valence bands ($V$) [40]. The properties of deformation potential tensor $\Xi_{ij}^{C,V}$ are determined by the crystalline symmetry of the material and the positions of the bottom of conduction band and the top of the valence band in the Brillouin zone [37-42].

Neglecting the strain-induced changes in the density of states (DOS) in the energy bands, one can express the impact on the strain of the equilibrium concentration of the electrons in the conduction and holes in the valence bands in terms of this ways introduced deformation potential [74, 75]:

$$\begin{aligned} n_C(\mathbf{r}) &= \int_{-\infty}^{\infty} \left[ 1 + \exp\left( \frac{\varepsilon + E_C + \Xi_{ij}^C u_{ij}(\mathbf{r}) - E_F - q\varphi(\mathbf{r})}{k_B T} \right) \right]^{-1} g_C(\varepsilon) d\varepsilon \\ &\approx n_{C0} \exp\left( \frac{-\Xi_{ij}^C u_{ij}(\mathbf{r}) + q\varphi(\mathbf{r})}{k_B T} \right) \end{aligned}, \qquad (8a)$$

$$\begin{aligned} p(\mathbf{r}) &= \int_{-\infty}^{\infty} \left[ 1 + \exp\left( -\frac{\varepsilon + E_V - \Xi_{ij}^V u_{ij}(\mathbf{r}) - E_F - q\varphi(\mathbf{r})}{k_B T} \right) \right]^{-1} g_V(\varepsilon) d\varepsilon \\ &\approx p_0 \exp\left( \frac{-\Xi_{ij}^V u_{ij}(\mathbf{r}) - q\varphi(\mathbf{r})}{k_B T} \right) \end{aligned}, \qquad (8b)$$

where $k_B$=1.3807×10$^{-23}$ J/K, $T$ is the absolute temperature $E_F$ is the Fermi level; $q$ is the absolute value of electron charge. Functions $g_m(x)$ with the script $m = C,V$ are the densities of states (DOS). [76]

Approximate equalities in Eq.(8) correspond to the Boltzmann-Planck-Nernst approximation that is widely used for MIECs (see e.g. Riess et al papers [77, 78, 79]). In this approximation, in the absence of external potential and strains the equilibrium concentrations of the electrons in conduction band and holes in the valence band, $n_{C0}$ and $p_0$, read $n_{C0} = \int_{-\infty}^{\infty} d\varepsilon \cdot g_C(\varepsilon) \exp\left( \frac{-E_C + E_F - \varepsilon}{k_B T} \right)$ and $p_0 = \int_{-\infty}^{\infty} d\varepsilon \cdot g_V(\varepsilon) \exp\left( \frac{E_V - E_F + \varepsilon}{k_B T} \right)$, respectively.

One readily shows that a converse effect to that discussed above (i.e. the stress/strain produced by the carrier redistribution), conditioned by the deformation potential, should exist, namely:

$$\sigma_{ij}(\mathbf{r}) = c_{ijkl} u_{kl}(\mathbf{r}) + \Xi_{ij}^C (n_C(\mathbf{r}) - n_{C0}) + \Xi_{ij}^V (p(\mathbf{r}) - p_0), \qquad (9a)$$

$$u_{ij}(\mathbf{r}) = s_{ijkl} \sigma_{kl}(\mathbf{r}) - \tilde{\Xi}_{ij}^C (n_C(\mathbf{r}) - n_{C0}) - \tilde{\Xi}_{ij}^V (p(\mathbf{r}) - p_0). \qquad (9b)$$

The deformation potential tensors in Eq.(9a) and (9b) are related as $\tilde{\Xi}_{ij}^{C,V} = s_{ijkl} \Xi_{kl}^{C,V}$.

Let us demonstrate the validity of Eq.(9a) for the electrons in the conductive band, obeying the classical statistics. We start from the expression for the free energy density of electrons in conductive band [74]:

$$\frac{F}{V}=\frac{1}{V}\sum_{i}\left[f_i\left(E_C\left(u_{ij}\right)+\varepsilon_i\right)+k_BT\left(f_i\ln f_i-f_i\right)\right]. \tag{10a}$$

Here $f_i=\exp\left(-\frac{E_C\left(u_{ij}\right)+\varepsilon_i-E_F-q\varphi}{k_BT}\right)$ is the probability of the occupation of the $i$-th state in the band by an electron, the summation is performed over conduction band, and $V$ is the system volume. Alternatively, $f_i$ can be expressed in terms of the density of the electrons, $n_C=\frac{1}{V}\sum_i f_i\equiv N_C\exp\left(-\frac{E_C\left(u_{ij}\right)}{k_BT}\right)$, and the density of state, $N_C$, in the conductive band, namely

$$f_i=\frac{n_C}{N_C}\exp\left(-\frac{E_C\left(u_{ij}\right)}{k_BT}\right). \tag{10b}$$

Combining (10a) and (10b), the free energy density can be expressed in term of its independent variables $u_{ij}$, $n_C$, and $T$:

$$\frac{F}{V}=n_CE_C\left(u_{ij}\right)+k_BTn_C\left(\ln\left(\frac{n_C}{N_C}\right)-1\right). \tag{10c}$$

By definition

$$\sigma_{ij}=\frac{\partial}{\partial u_{ij}}\left(\frac{F}{V}\right)\Bigg|_{T,n_C}=n_C\left(\frac{\partial}{\partial u_{ij}}E_C\left(u_{ij}\right)-\frac{k_BT}{N_C}\frac{\partial N_C}{\partial u_{ij}}\right)=n_C\Xi^C_{ij}-n_C\frac{k_BT}{N_C}\frac{\partial N_C}{\partial u_{ij}}\approx n_C\Xi^C_{ij}. \tag{11}$$

Thus, neglecting the strain dependence of the density of states and keeping in mind that we are interested in the strain difference between the initial state of the system and that with a changed electron density, we arrive at the second r.h.s. term from Eq.(9a). The calculations for the stress induced by the variation of the holes density are similar. The impact of the last term, $n_C\frac{k_BT}{N_C}\frac{\partial N_C}{\partial u_{ij}}$, appeared small for semiconductors, since the strain dependence of the effective mass is typically much smaller than the band gap dependence determined by deformation potential (see e.g. Ref. [80]).

### 3. Elastic fields: flexoelectric, Vegard and electron-phonon contributions

The total stress contains flexoelectric contribution in accordance with Eq.(4), Vegard contribution in accordance with Eq.(5) and electron-phonon contribution in accordance with Eq.(9). Thus, the strain and stress tensors are related as:

$$\sigma_{ij}(\mathbf{r}) = c_{ijkl}u_{kl}(\mathbf{r}) + \begin{pmatrix} \Xi_{ij}^{C}(n_C(\mathbf{r}) - n_{C0}) + \Xi_{ij}^{V}(p(\mathbf{r}) - p_0) + \\ -\beta_{ij}^{a}(N_a^{-}(\mathbf{r}) - N_{a0}^{-}) - \beta_{ij}^{d}(N_d^{+}(\mathbf{r}) - N_{d0}^{+}) \end{pmatrix} - \gamma_{ijkl}\frac{\partial^2\varphi}{\partial x_k \partial x_l}. \quad (12a)$$

The strain tensor can be expressed via the stress tensor (10) as:

$$u_{ij}(\mathbf{r}) = s_{ijkl}\sigma_{kl}(\mathbf{r}) + \begin{pmatrix} \tilde{\beta}_{ij}^{a}(N_a^{-}(\mathbf{r}) - N_{a0}^{-}) + \tilde{\beta}_{ij}^{d}(N_d^{+}(\mathbf{r}) - N_{d0}^{+}) \\ -\tilde{\Xi}_{ij}^{C}(n_C(\mathbf{r}) - n_{C0}) - \tilde{\Xi}_{ij}^{V}(p(\mathbf{r}) - p_0) \end{pmatrix} + \tilde{\gamma}_{ijkl}\frac{\partial^2\varphi(\mathbf{r})}{\partial x_k \partial x_l}. \quad (12b)$$

The inverse effects tensors and flexoelectric coefficients in Eq.(11b) are introduced as

$$\tilde{\Xi}_{ij}^{C,V} = s_{ijkl}\Xi_{kl}^{C,V}, \quad \tilde{\beta}_{ij}^{a,d} = s_{ijkl}\beta_{kl}^{a,d}, \qquad \tilde{\gamma}_{ijkl} = s_{ijmn}\gamma_{mnkl}. \quad (13a)$$

Note, that Eqs.(12) require the reference lattice determination. The reference lattice is regarded strain-free for the case of zero electric potential: $\varphi = 0$ and therefore $n_C(\mathbf{r}) = n_{C0}$, $p(\mathbf{r}) = p_0$, $N_a^{-}(\mathbf{r}) = N_{a0}^{-}$, $N_d^{+}(\mathbf{r}) = N_{d0}^{+}$.

Considering the case of isotropic media, for which $\Xi_{ij}^{C,V} = \Xi^{C,V}\delta_{ij}$, $\beta_{ij}^{a,d} = \beta^{a,d}\delta_{ij}$ and $\gamma_{ijkl} = \gamma_D\delta_{ij}\delta_{kl} + \gamma_S(\delta_{ik}\delta_{jl} + \delta_{il}\delta_{jk})$, in Voigt notations Eq (13a) can be simplified as

$$\begin{gathered} \tilde{\Xi}_{ij}^{C,V} = \Xi^{C,V}(s_{11} + 2s_{12})\delta_{ij}, \quad \tilde{\beta}_{ij}^{a,d} = \beta^{a,d}(s_{11} + 2s_{12})\delta_{ij}, \\ \tilde{\gamma}_{33} = \tilde{\gamma}_{22} = \tilde{\gamma}_{11} = s_{11}\gamma_{11} + 2s_{12}\gamma_{12}, \quad \tilde{\gamma}_{12} = \gamma_{11}s_{12} + \gamma_{12}(s_{11} + s_{12}), \quad \tilde{\gamma}_{44} = \gamma_{44}s_{44}. \end{gathered} \quad (13b)$$

Note that the group of *k* at the Γ point in the Brillouin zone is isomorphic to the point group of the lattice so the Γ point has full crystal symmetry. The Γ point symmetry determines the dilatational deformation potential tensor [40]. Thus non-diagonal components of dilatational deformation potential tensor as well as of the Vegard strain tensor are possible only for monoclinic and triclinic symmetry materials (since these tensors are symmetric polar ones, their symmetry properties are the same as for e.g. dielectric susceptibility tensors, see e.g. Ref. [81]).

Estimation of the deformation potential tensor trace performed in the Tomas-Fermi approximation [37] yields the magnitude of $\beta \sim 1$ eV and $\tilde{\beta} \sim 10^{-30}$ m$^3$ for Li-containing ionics [23, 25, 82]. Unfortunately, the Tomas-Fermi approximation can significantly underestimate the deformation tensor value for oxide semiconductor materials and metal-insulators with charge gap up to the order of magnitude [37, 73]. Experimental values are not available, albeit are probably accessible for density-functional type modelling. In comparison, for Si- or Ge-based semiconductors experimental values are $\tilde{\Xi}$ ~5 – 10 eV and $\tilde{\Xi}$ ~(1 – 5) $10^{-30}$m$^3$ [40, 83]. Using the values and typical range of concentration variations, namely:(a) 1% deviation from stoichiometric concentration $10^{28}$ m$^{-3}$ for ions gives $\left(N_a^-(\mathbf{r})-N_{a0}^-\right)\sim 10^{26}$ m$^{-3}$; (b) 10 – 100% deviation of electrons and holes concentration in the regions of the depletion/accumulation regions is about $(p_0-p(\mathbf{r}))\sim 10^{27}\,\text{m}^{-3}$, we estimate that the contributions of Vegard effect $\tilde{\beta}_{ij}^a\left(N_a^-(\mathbf{r})-N_{a0}^-\right)$ and deformation potential $\tilde{\Xi}_{ij}^V(p_0-p(\mathbf{r}))$ in Eq.(11) are comparable for ionics.

## 4. The strain-voltage response in decoupling approximation

Here, we illustrate the contribution of ions and electrons migration in the applied electric field to the strain response of the MIEC surface. It is seen from Eqs.(12) that the Vegard expansion, deformation potential and flexoelectric effect couple the stress field with the carriers distribution, requiring the solution of fully coupled problem. However, in the most cases the changes of band structure due to the external pressure is rather weak (e.g., for Ge band gap changes only on about 1% for rather high strain of about $10^{-3}$ [38]). Hence, when calculating the space charges distributions the stress contribution can be neglected in the first approximation. Then the ionic and electrostatic field distributions are substituted in Eqs.(12) to yield mechanical responses. The approach is the decoupling approximation to account for the effects of deformation potential, chemical expansion and flexoelectric effect.

### *4.1. Electrochemical Strain Microscopy of the MIEC*

Both ionic and electronic contributions to the local strain can be measured and distinguished by the Electrochemical Strain Microscopy (ESM) [11, 12, 13, 14, 84]. For the ionically blocking tip electrode, the electron transfer between the tip and the surface and non-uniform electrostatic field result in mobile ions and electrons redistribution within the solid, but no electrochemical process at the interface occurs [12]. The schematic of the system is shown in **Fig. 1a.**

Lame-type equation for the mechanical displacement $u_i$ can be obtained from the equation of mechanical equilibrium $\partial\sigma_{ij}(\mathbf{r})/\partial x_i = 0$, where the stress tensor $\sigma_{ij}(\mathbf{r})$ is given by Eq.(11a), namely:

$$c_{ijkl}\frac{\partial^2 u_k}{\partial x_j \partial x_l} = -\frac{\partial}{\partial x_j}\begin{pmatrix} -\beta_{ij}^a\left(N_a^-(\mathbf{r}) - N_{a0}^-\right) - \beta_{ij}^d\left(N_d^+(\mathbf{r}) - N_{d0}^+\right) \\ -\gamma_{ijkl}\frac{\partial^2\varphi}{\partial x_k \partial x_l} + \Xi_{ij}^C\left(n_C(\mathbf{r}) - n_{C0}\right) + \Xi_{ij}^V\left(p(\mathbf{r}) - p_0\right) \end{pmatrix} \qquad (14)$$

Mechanical boundary conditions [85] corresponding to the ESM experiments [11] are defined on the mechanically free interface, $z = 0$, where the normal stress $\sigma_{3i}$ is absent, and on clamped interface $z = h$, where the displacement $u_i$ is fixed:

$$\sigma_{3i}(x_1, x_2, z = 0) = 0, \qquad u_i(x_1, x_2, z = h) = 0. \qquad (15)$$

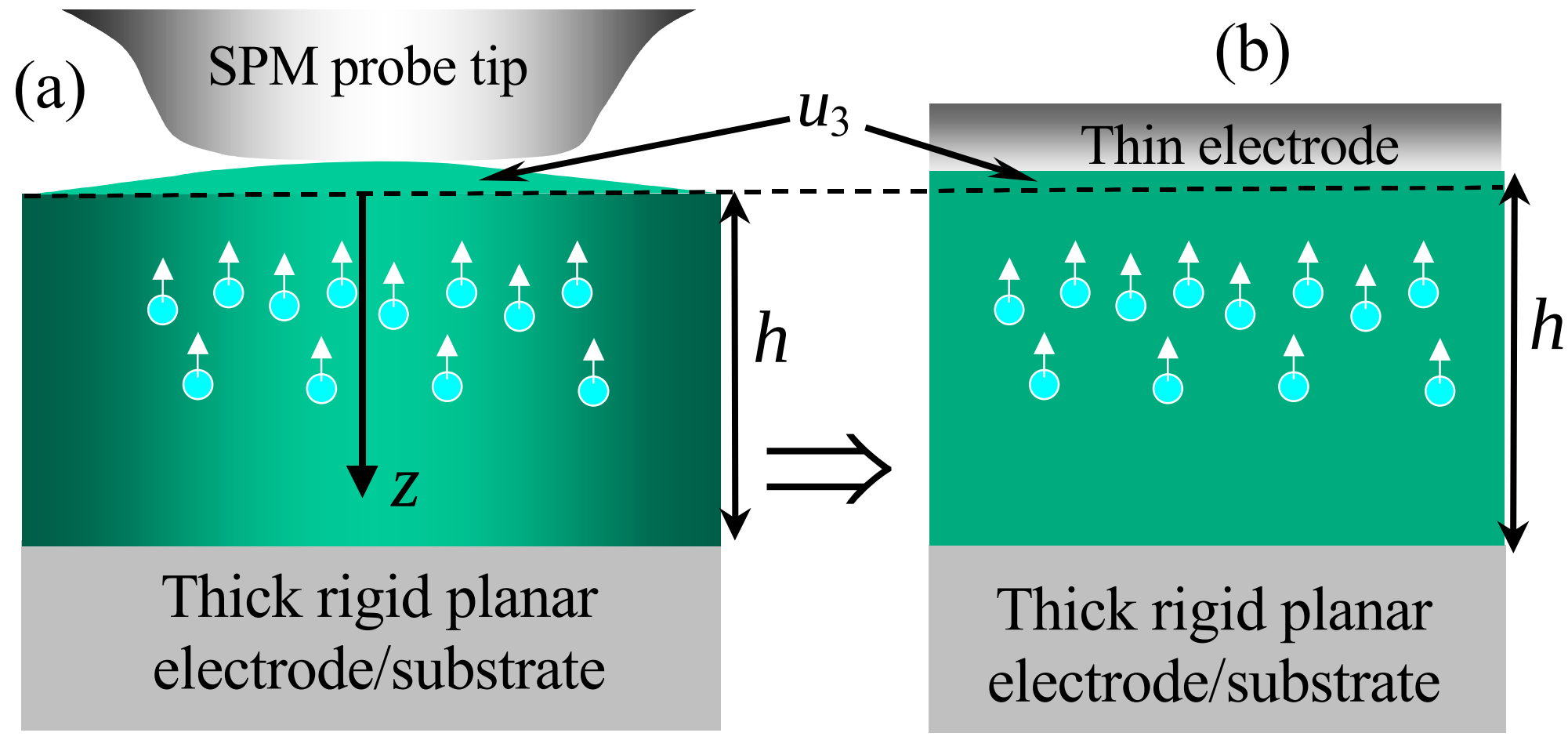

**Fig. 1.** Schematics of ESM measurements with a flattened SPM tip (a) is approximated by the (b) strain response of the 1D-system, where $u_3$ is the surface displacement for fixed back interface. Voltage $V_0$ is applied to the top electrode.

The tip bias – induced displacement of the MIEC surface at the point $x_3$=0, i.e. surface displacement at the tip-surface junction detected by SPM electronics, for elastically isotropic semi-space can be calculated in decoupling approximation [12], using the appropriate tensorial Green function for elastic semi-space (listed in e.g. Ref.[86]) or thin film (derived in Refs.[87, 88]). Decoupling approximation regards the flexoelectric effect and strain contribution small enough not to perturb the electrostatic potential and carrier distributions in the first approximation. Thus below we determine the electric potential from the Eq.(2) with carriers distribution (6) and (8) without strain terms and then substitute the potential and carriers distribution into Eq.(14).

Note, that decoupling approximation introduced earlier for PFM [89, 90], are sufficiently rigorous for materials with low electromechanical coupling coefficients, i.e. for all non-piezoelectrics considered in the paper. The accuracy of the decoupling approximation is proportional to the square of the electromechanical coupling coefficients, which generally does not exceed $10^{-2}$ for non-ferroelectrics.

### *4.2. Strain response of the surface layers*

The schematic of the capacitor-like structure that models a disc-like SPM tip is illustrated in **Fig. 1b.** We consider a MIEC film of thickness, $h$, sandwiched between the planar electrodes. For the strain measurements, the top electrode is considered to be mechanically free (e.g. ultra-thin, or liquid, or soft polymer), so that its motion does not affect significantly the mechanical displacement of the MIEC film surface. Voltage $V_0$ is applied to the top electrode, the bottom electrode is earthed:

$$\varphi(z)=V_0\approx const,\quad \varphi(h)=0\,. \tag{16}$$

The voltage drop between the top and bottom electrode causes the 1D-redistribution of the carrier concentration in z-direction.

Using Eq.(1) from Ref.[11], equilibrium mechanical displacement of the MIEC surface caused by the flexoelectric, electronic and ionic contributions can be calculated as:

$$u_3(z=0)=-\int_0^h dz\left(\begin{array}{l}\left(\tilde{\Xi}^C_{33}-\dfrac{2s_{12}\tilde{\Xi}^C_{11}}{s_{11}+s_{12}}\right)\left(n_C(z)-n_{C0}\right)+\left(\tilde{\Xi}^V_{33}-\dfrac{2s_{12}\tilde{\Xi}^V_{11}}{s_{11}+s_{12}}\right)\left(p(z)-p_0\right)\\ +\left(\tilde{\beta}^a_{33}-\dfrac{2s_{12}\tilde{\beta}^a_{11}}{s_{11}+s_{12}}\right)\left(N^-_{a0}-N^-_a(z)\right)+\left(\tilde{\beta}^d_{33}-\dfrac{2s_{12}\tilde{\beta}^d_{11}}{s_{11}+s_{12}}\right)\left(N^+_{d0}-N^+_d(z)\right)\\ +\left(\tilde{\gamma}_{3333}-\dfrac{2s_{12}\tilde{\gamma}_{1133}}{s_{11}+s_{12}}\right)\dfrac{d^2\varphi}{dz^2}\end{array}\right). \tag{17}$$

Note that the contribution of the electron-phonon coupling (first two terms in Eq.(17)) as well as the flexoelectric effect (the last term) into the local surface displacement can be comparable with the first terms originated from the chemical expansion. Moreover, using the order of magnitude estimate of $\gamma\sim1\ 10^{-10}$ C/m, the flexoelectric contribution to the PFM signal is about 12 pm/V. Note that for biological systems the flexoelectric response can be significantly stronger than the piezoelectric one [91], since they are elastically soft.

Using the decoupling approximation in the 1D-Poisson equation, $\varepsilon_0\varepsilon_{33}\frac{d^2\varphi(\mathbf{r})}{dz^2}=-q\left(p-p_0+n_{C0}-n_C+N_{a0}^- -N_a^- +N_d^+ -N_{d0}^+\right)$, i.e. neglecting here the flexoelectric term $\gamma_{ij33}\frac{d^2u_{ij}}{dz^2}$, and regarding that $\left(-p_0+n_{C0}+N_{a0}^- -N_{d0}^+\right)=0$ due to the electroneutrality in the bulk MIEC, Eq.(17) can be simplified as

$$u_3(z=0)\approx-\int_0^h dz\begin{pmatrix}\lambda\left(\tilde{\Xi}^C,\tilde{\gamma}\right)\left(n_C(z)-n_{C0}\right)+\lambda\left(\tilde{\Xi}^V,-\tilde{\gamma}\right)\left(p(z)-p_0\right)+\\ +\mu\left(\tilde{\beta}^a,\tilde{\gamma}\right)\left(N_a^-(z)-N_{a0}^-\right)+\mu\left(\tilde{\beta}^d,-\tilde{\gamma}\right)\left(N_d^+(z)-N_{d0}^+\right)\end{pmatrix} \tag{18}$$

It is seen from Eq.(18) that the MIEC surface displacement is proportional ***to the total charge of each species.*** Thus only the injected charges control the displacement. Note, that the relation between the total charge and electrostatic potential on the semiconductor surface are well established [74].

In Eq.(18) we introduced the designations for the flexo-electro-chemical coupling constants as

$$\lambda\left(\tilde{\Xi},\tilde{\gamma}\right)=\tilde{\Xi}_{33}-\frac{2s_{12}\tilde{\Xi}_{11}}{s_{11}+s_{12}}+\left(\tilde{\gamma}_{3333}-\frac{2s_{12}\tilde{\gamma}_{1122}}{s_{11}+s_{12}}\right)\frac{q}{\varepsilon_0\varepsilon_{33}}, \tag{19}$$

$$\mu\left(\tilde{\beta},\tilde{\gamma}\right)=-\tilde{\beta}_{33}+\frac{2s_{12}\tilde{\beta}_{11}}{s_{11}+s_{12}}+\left(\tilde{\gamma}_{3333}-\frac{2s_{12}\tilde{\gamma}_{1122}}{s_{11}+s_{12}}\right)\frac{q}{\varepsilon_0\varepsilon_{33}}, \tag{20}$$

where the first terms originated from the deformation potential or Vegard tensors, while the last ones originated from the flexoelectric coupling.

Flexoelectric effect contribution into the coupling constants λ and μ from Eqs.(19)-(20) is estimated in the **Table 1**. It is seen from the **Table 1** that the flexoelectric contribution ranges from 0.1 to 10 eV for crystalline dielectrics, that is comparable to or much higher than the chemical expansion and deformation potential contributions, which are ~0.5 – 5 eV for ionics. For incipient ($SrTiO_3$) and normal ($Pb(Zr,Ti)O_3$ and $BaTiO_3$) ferroelectrics the flexoelectric effect contribution is much higher than the other ones.

**Table 1. Flexoelectric effect contribution into the coupling constants λ and μ**

| Material | Flexo-electric tensor γ (nC/m) | ε (at 300 K) | Flexoelectric coupling constant (eV) $\left(\gamma_{33}-\frac{2s_{12}\gamma_{12}}{s_{11}+s_{12}}\right)\frac{q}{\varepsilon_0\varepsilon_{33}}$ | Flexoelectric coupling constant (m$^3$) $\left(\tilde{\gamma}_{33}-\frac{2s_{12}\tilde{\gamma}_{12}}{s_{11}+s_{12}}\right)\frac{q}{\varepsilon_0\varepsilon_{33}}$ | Ref. |
|---|---|---|---|---|---|
| crystalline dielectrics, elastomers | ~0.01–0.1 | ~10 | ~0.1–1 | ~(0.1–1) $10^{-30}$ | [92] |
| single crystal $SrTiO_3$ | $\gamma_{3333}=-9$, $\gamma_{1122}=4$, $\gamma_{1212}=3$ | 300 | −2 | −1.7 $10^{-30}$ | [54] |
| ceramic PZT-5H | $\gamma_{1122}=500$ | 2200 | ~30 | ~5 $10^{-29}$ | [51] |
| ceramic $BaTiO_3$ | $\gamma_{1122}=10^4$ (with domain walls) | 2000 | ~500 | ~ $10^{-27}$ | [52] |
| single crystal $BaTiO_3$ | $\gamma_{3333}=-0.37$ *ab initio* at 0 K | 200 | ~0.5 | ~ $10^{-29}$ | [93] |

For numerical estimations, we consider the situation when the MIEC film with mobile acceptors and holes is at the thermodynamic equilibrium (i.e. all currents are absent). The analytical solution for acceptors and holes redistribution in a thick MIEC film and its surface displacement are derived in **Appendix A** assuming that film thickness $h>>R_S$, where the screening radius $R_S=\sqrt{\frac{\varepsilon_{33}\,\varepsilon_0 k_B T}{2p_0 q^2}}$.

Substitution of the total charge of each species in Eq.(18) in the limit $h>>R_S$ gives the estimations for the MIEC surface displacement. Note, that for the ionically blocking planar top and substrate electrodes the identity $\int_0^h dz\left(N_a^-(z,t)-N_{a0}^-\right)=0$ is valid [77, 78, 79, 94], since the total amount of ionized acceptors is conserved. Thus only the electron subsystem contributes to the surface displacement (18) for the ion-blocking electrodes as:

$$u_3(V_0)\approx\lambda\left(\tilde{\Xi}^V,-\tilde{\gamma}\right)\sqrt{\frac{2\varepsilon_{33}\,\varepsilon_0 k_B T}{q^2}N_{a0}^-}\left(1-\exp\left(-\frac{qV_0}{2k_B T}\right)\right),\quad h>>R_S, \tag{21a}$$

$$u_3(V_0) \approx \lambda\left(\tilde{\Xi}^V,-\tilde{\gamma}\right)\sqrt{\frac{\varepsilon_{33}\,\varepsilon_0 k_B T}{2q^2}N_{a0}^-}\,\frac{qV_0}{k_B T},\quad h >> R_S,\quad \left|qV_0\right| << k_B T\,. \qquad (21b)$$

It follows from Eq.(21b) that in the linear approximation the electronic surface displacement is proportional to the applied voltage $V_0$, stoichiometric acceptor concentration $N_{a0}^-$, tensorial deformation potential $\tilde{\Xi}_{ii}^V$ and flexoelectric effect $\tilde{\gamma}_{iijj}$ via the coupling constant $\lambda\left(\tilde{\Xi}^V,\tilde{\gamma}\right)$.

Correspondingly, even though strain contribution can be neglected when considering the chemical potentials and carrier distribution for a film with ion-blocking interfaces, we could not neglect deformation potential and flexoelectric effect influence on elastic subsystem, since it is the only source of strain in the case. The measurements of the MIEC surface displacement placed between thin ionically blocking planar electrodes can be performed by the interferometer.

For ionically conducting electrode(s) substitution of the total charge of each species in Eq.(18), yield the mixed ionic-electronic strain-voltage response as:

$$u_3(V_0) \approx -\left(\begin{array}{l}\lambda\left(\tilde{\Xi}^V,-\tilde{\gamma}\right)\sqrt{\frac{2\varepsilon_{33}\varepsilon_0 k_B T}{q^2}N_{a0}^-}\left(\exp\left(-\frac{qV_0}{2k_B T}\right)-1\right)\\ +\mu\left(\tilde{\beta}^a,\tilde{\gamma}\right)\sqrt{\frac{2\varepsilon_{33}\varepsilon_0 k_B T}{q^2}N_{a0}^-}\left(\exp\left(\frac{qV_0}{2k_B T}\right)-1\right)\end{array}\right). \qquad (22)$$

Equation (22) is derived for thick films, $h >> R_S$. It is seen from Eq.(22) that in the linear approximation the mixed ionic-electronic surface displacement is proportional to the applied voltage $V_0$, acceptors stoichiometry concentration $N_{a0}^-$, deformation tensors $\tilde{\Xi}_{ii}^V$, Vegard expansion tensors $\tilde{\beta}_{ii}^a$ and flexoelectric coefficients $\tilde{\gamma}_{iijj}$ via the coupling constants $\lambda\left(\tilde{\Xi}^V,-\tilde{\gamma}\right)$ and $\mu\left(\tilde{\beta}^a,\tilde{\gamma}\right)$.

Note, that realistic ESM tip is nano- or submicro-sized. Therefore the possibility of the ions motion in lateral direction rather leads to the condition of ion-conducting tip electrode than ion-blocking.

Electronic strain-voltage response $u_3(V_0)$ of the MIEC film placed between ionically-blocking electrodes as calculated from Eq.(21) is shown in **Figs. 2a,b**. The electronic strain-voltage response demonstrate strong asymmetry ("diode-type rectification") with the change of electric voltage polarity: for positive $V_0 > 0$ strong saturation occurs at very small response values, while for negative $V_0 < 0$ the response rapidly increases linearly and reaches noticeable values $u_3(V_0)$ ~1-10 nm at $V_0 \sim 1$ V. Probably, non-linear behavior should be reached for negative voltages in practice since the hole statistics eventually becomes degenerated in the case of strong depletion/accumulation near the MIEC surface; but the effect of carrier degeneration is beyond the approximation (21). The response absolute value $u_3(V_0)$ decreases as the ions concentration decrease (follow arrow direction for the typical values of mobile acceptor concentration $N_{a0}^{-}$=$10^{23}$ – $10^{26}$ m$^{-3}$ in the **Figs. 2a,b**).

Mixed ionic-electronic strain-voltage response $u_3(V_0)$ of the MIEC film placed between the electrodes, one or both of which is ionically-conducting, was calculated from Eq.(22) and are shown in **Figs. 2c,d**. In logarithmic voltage scale the asymmetry appearing with the change of electric voltage polarity is rather weak. However, it becomes obvious on the linear scale (compare **Figs. 2c** and **d**). The effect originates from the fact that the typical electronic contribution $\tilde{\Xi}_{ii}^{V}$ ~$10^{-31}$m$^3$ is only one order of magnitude smaller than the ionic, $\tilde{\beta}_{ii}^{a}$ ~$10^{-30}$m$^3$.

In dimensionless units the strain-voltage response depends on one parameter $qV_0/(k_B T)$, as anticipated from the diode-theory for the case $h >> R_S$ (see **Figs. 2b** and **d)**.

The crossover from the dominantly ionic ($\left|\lambda\left(\tilde{\Xi}^{V},-\tilde{\gamma}\right)\right| << \left|\mu\left(\tilde{\beta}^{a},\tilde{\gamma}\right)\right|$) to electronic ($\left|\lambda\left(\tilde{\Xi}^{V},-\tilde{\gamma}\right)\right| >> \left|\mu\left(\tilde{\beta}^{a},\tilde{\gamma}\right)\right|$) strain-voltage response is shown in **Figs. 3.** In the case $\left|\lambda\left(\tilde{\Xi}^{V},-\tilde{\gamma}\right)\right| = \left|\mu\left(\tilde{\beta}^{a},\tilde{\gamma}\right)\right|$ the strain-voltage curve is symmetric.

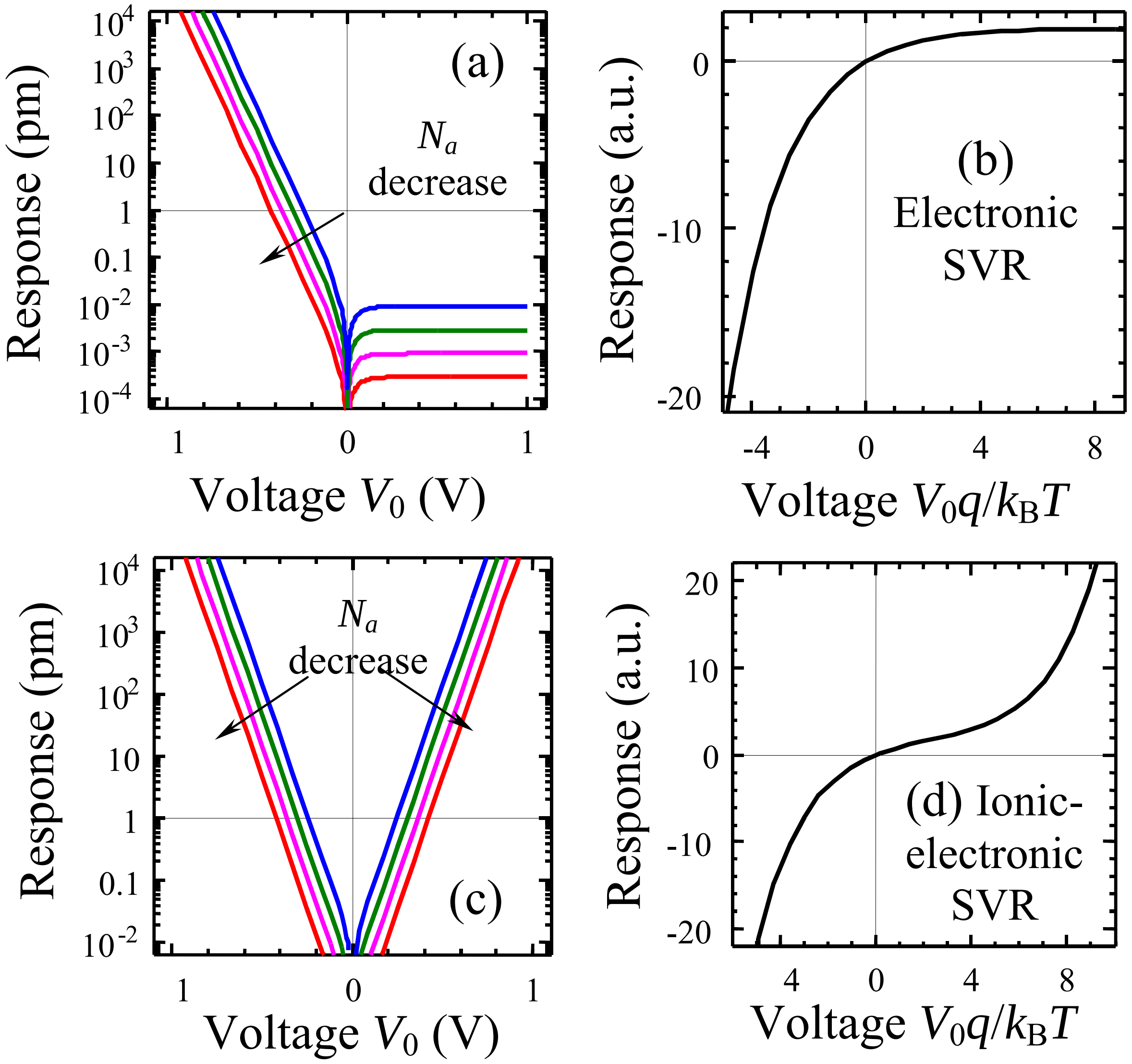

**Fig. 2.** (a, b) Electronic strain-voltage response (absolute value SVR) $u_3(V_0)$ of the MIEC film placed between ionically-blocking electrodes. (c, d) Mixed ionic-electronic strain-voltage response $u_3(V_0)$ of the MIEC film placed between ionically-blocking top electrode and ionically-conducting bottom electrode calculated for different values of mobile acceptor concentration $N_{a0}^{-}$=$10^{23}$, $10^{24}$, $10^{25}$, $10^{26}$ m$^{-3}$ (arrow near the curves), room temperature $T$=300 K, coupling constant $\lambda\left(\tilde{\Xi}^{V},-\tilde{\gamma}\right)=10^{-31}$m$^3$, $\mu\left(\tilde{\beta}^{a},\tilde{\gamma}\right)=10^{-30}$m$^3$, MIEC film thickness $h=100R_S$. Plots (b, d) are in dimensionless units.

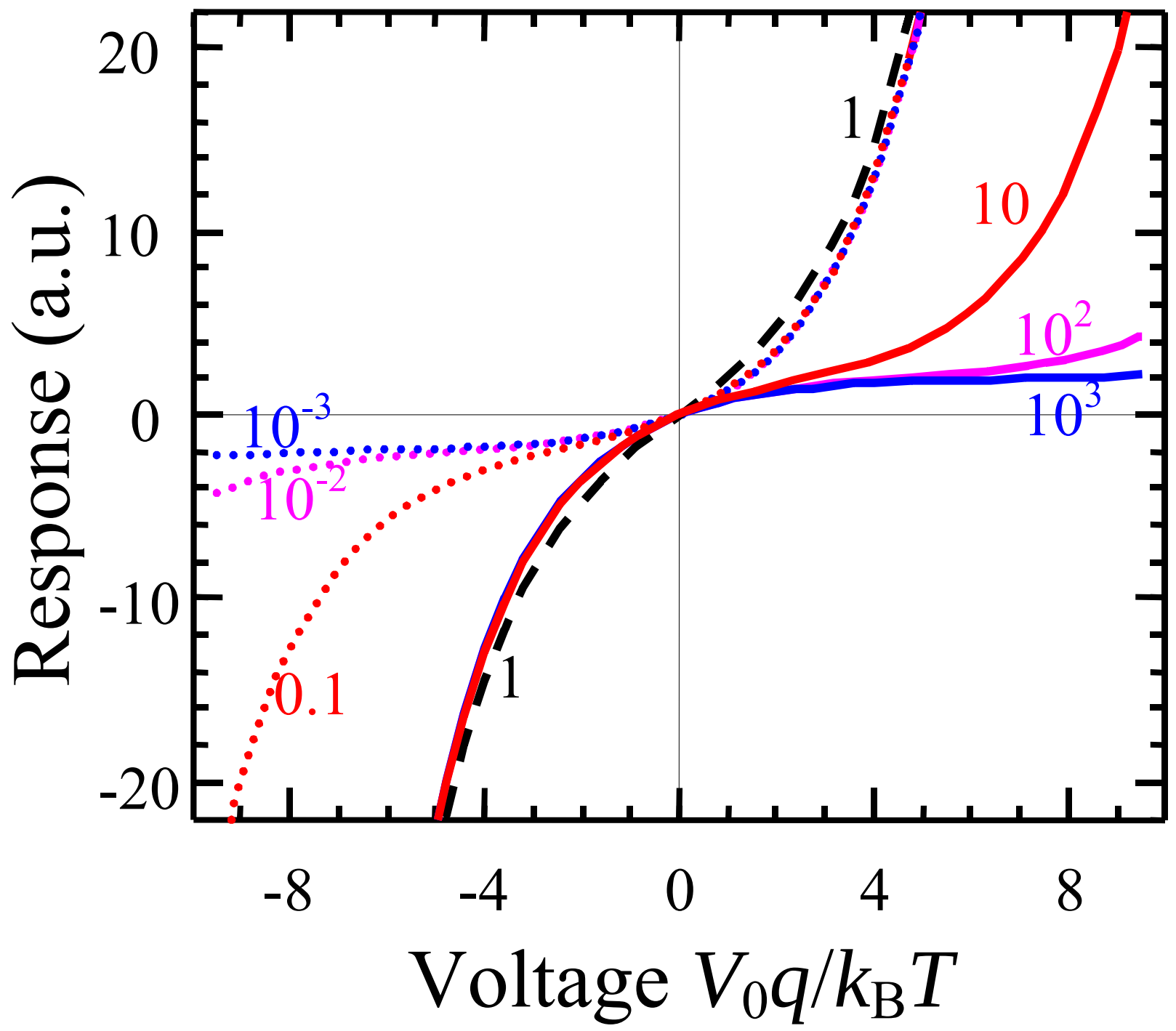

**Figs. 3.** The crossover between from the dominantly ionic to electronic strain-voltage response: $\left|\lambda\left(\tilde{\Xi}^V,-\tilde{\gamma}\right)/\mu\left(\tilde{\beta}^a,\tilde{\gamma}\right)\right|=0.001$, 0.01, 0.1, 1, 10, 100, 1000 (figures near the curves) Acceptor concentration $N_{a0}^{-}=10^{24}$ m$^{-3}$, other parameters are the same as in **Fig. 2.**

### 5. Summary

We derive the generalized form of the bias-strain-concentration equation describing the linear relation between the concentration of diffusing species, flexoelectric and electronic effects in mixed ionic-electronic conductors. The estimates of the electronic and ionic contributions into the stain-voltage response of the mixed ionic-electronic conductors show that they are of the same order, and hence one could not neglect the electronic contribution into the surface displacement of the sample with ion-blocking interfaces (injection from the tip). To the best of our knowledge the contribution of the electron-phonon and flexoelectric coupling into the local surface displacement of the mixed ionic-electronic has not been previously discussed. Evolved approach can be extended to treat electrochemically induced mechanical phenomena in solid state ionics towards analytical theory and phase-field modeling of mixed ionic-electronic conductors.

## Acknowledgements

Authors are grateful to V. Kharton (U. Aveiro) for useful discussions and valuable advice. ANM, EEA and LQC research sponsored by in part by Ukraine State Committee on Science, Innovation and Information (UU30/004) and National Science Foundation (DMR-0908718 and DMR-0820404). ANM, EEA and SVK acknowledge user agreement with CNMS N UR-08-869. This research was supported in part (SVK) at the Center for Nanophase Materials Sciences, which is sponsored at Oak Ridge National Laboratory by the Division of Scientific User Facilities, U.S. Department of Energy. AKT acknowledges Swiss National Science Foundation for financial support.

## Appendix A. Equilibrium distribution of the potential and space charge in a semi-infinite MIEC (decoupling approximation)

Equilibrium state corresponds to the absence of ionic (acceptor, donor) and electronic (hole) currents. In the linear drift-diffusion model the acceptor $J_a$ and hole $J_p$ currents have the form

$$\begin{cases} J_a = -\left(D_a \dfrac{d}{dz} N_a^- - \eta_a N_a^- \dfrac{d}{dz}\varphi\right) = 0, \\ J_p = -\left(D_p \dfrac{d}{dz} p + \eta_p p \dfrac{d}{dz}\varphi\right) = 0 \end{cases} \tag{A.1}$$

Hereinafter we regard that the diffusion coefficients $D_{a,p}$ and mobilities $\eta_{a,p}$ obey the Nerst-Einstein relation $\eta_d / D_d = \eta_n / D_n = q/(k_B T)$, where $k_B$=1.3807×10$^{-23}$ J/K, $T$ is the absolute temperature.

The solution of Eqs.(A.1) is

$$N_a^-(z) = N_0 \exp\left(\frac{q\varphi(z)}{k_B T}\right), \tag{A.2a}$$

$$p(z) = p_0 \exp\left(-\frac{q\varphi(z)}{k_B T}\right) \tag{A.2b}$$

Note, that solutions (A.2) coincide with Eqs.(6b) and (8b) as anticipated. Using the decoupling approximation (i.e. neglecting here the term $\gamma_{ij33}\, d^2u_{ij}/dz^2$ ), the boundary problem for electrostatic potential distribution in the following form:

$$\begin{cases} \dfrac{d^2\varphi(z)}{dz^2} = -\dfrac{q}{\varepsilon_0\varepsilon}\left(p_0 \exp\left(-\dfrac{q\varphi(z)}{k_B T}\right) - N_0 \exp\left(\dfrac{q\varphi(z)}{k_B T}\right)\right), \\ \varphi(0) = V_0, \quad \varphi(h \to \infty) = 0, \quad E_z = -\left.\dfrac{d\varphi}{dz}\right|_{h\to\infty} = 0. \end{cases} \tag{A.3}$$

The condition of the potential and electric field vanishing at the infinity leads to the local space charge vanishing that is valid under the condition $N_0 = p_0$. Then equation (A.3) acquires the form

$$\frac{d^2\varphi(z)}{dz^2} = \frac{2qp_0}{\varepsilon_0\varepsilon}\sinh\left(\frac{q\varphi(z)}{k_B T}\right) \tag{A.4}$$

and can be integrated in a straightforward way. Multiplying both sides of the equation by the potential gradient we calculated the first integral as $\left(\frac{d\varphi(z)}{dz}\right)^2 = \frac{4p_0k_BT}{\varepsilon_0\varepsilon}\left(\cosh\left(\frac{q\varphi(z)}{k_BT}\right) - a\right)$, where the constant $a = 1$ from the boundary conditions of electric field vanishing at the infinity. Using new variable $u = \cosh(q\varphi/(k_BT))$ one could rewrite (A.4) as

$$\varphi(z) = \frac{4k_BT}{q}\operatorname{arctanh}\left(\tanh\left(\frac{qV_0}{4k_BT}\right)\exp\left(-\frac{z}{R_S}\right)\right) \tag{A.5a}$$

$$N_a^-(z) = p_0 \exp\left(\frac{q\varphi(z)}{k_BT}\right), \qquad p(z) = p_0 \exp\left(-\frac{q\varphi(z)}{k_BT}\right) \tag{A.5b}$$

Here we introduced the screening radius $R_S = \sqrt{\varepsilon\varepsilon_0 k_BT/(2p_0q^2)}$.

Substitution of Eqs.(A.5) in Eq.(18) in the limit $h >> R_S$ gives the estimations for the MIEC surface displacement. Note, that for the ionically blocking planar top and substrate electrodes the identity $\int_0^h dz\left(N_a^-(z,t) - N_{a0}^-\right) = 0$ is valid [77, 78, 79, 94], since the total amount of ionized acceptors is conserved. The conditions $\int_0^h dz\left(N_a^-(z,t)-N_{a0}^-\right)=0$ and $N_0=p_0$ lead to the expression for $p_0=N_0=N_{a0}^-\left(\frac{1}{h}\int_0^h dz\exp\left(\frac{q\varphi(z)}{k_BT}\right)\right)^{-1}$, and thus for the ion-blocking planar electrodes only the electron subsystem contributes to the surface displacement (18) as:

$$\begin{aligned} u_3(z=0) &= \lambda\left(\tilde{\Xi}^V,-\tilde{\gamma}\right)p_0\int_0^h dz\left(1-\exp\left(-\frac{q\varphi(z)}{k_BT}\right)\right) \\ &\equiv \lambda\left(\tilde{\Xi}^V,-\tilde{\gamma}\right)hN_{a0}^-\int_0^h dz\left(1-e^{-\frac{q\varphi(z)}{k_BT}}\right)\left[\int_0^h dz e^{\frac{q\varphi(z)}{k_BT}}\right]^{-1} \end{aligned} . \qquad \text{(A.6)}$$

Under the condition of high film thickness, $h>>R_S$, Eq.(A.6) reduces to

$$u_3(V_0)\approx\lambda\left(\tilde{\Xi}^V,-\tilde{\gamma}\right)\sqrt{\frac{2\varepsilon_{33}\,\varepsilon_0 k_BT}{q^2}N_{a0}^-}\left(1-\exp\left(-\frac{qV_0}{2k_BT}\right)\right),\quad h>>R_S, \qquad \text{(A.7a)}$$

$$u_3(V_0)\approx\lambda\left(\tilde{\Xi}^V,-\tilde{\gamma}\right)\sqrt{\frac{\varepsilon_{33}\,\varepsilon_0 k_BT}{2q^2}N_{a0}^-}\frac{qV_0}{k_BT},\quad h>>R_S,\quad |qV_0|<<k_BT\,. \qquad \text{(A.7b)}$$

It follows from Eq.(A.7b) that in the linear approximation the electronic surface displacement is proportional to the applied voltage $V_0$, stoichiometric acceptor concentration $N_{a0}^-$, tensorial deformation potential $\tilde{\Xi}_{ii}^V$ and flexoelectric effect $\tilde{\gamma}_{iijj}$ via the coupling constant $\lambda\left(\tilde{\Xi}^V,\tilde{\gamma}\right)$.

For ionically conducting electrode(s) substitution of Eqs.(A.5) with $p_0=N_0=N_{a0}^-$ in Eq.(18), yield the mixed ionic-electronic strain-voltage response as:

$$u_3(z=0)=-\begin{pmatrix} \lambda\left(\tilde{\Xi}^V,-\tilde{\gamma}\right)N_{a0}^-\int_0^h dz\left(\exp\left(-\frac{q\varphi(z)}{k_BT}\right)-1\right)+ \\ +\lambda\left(\tilde{\beta}^a,\tilde{\gamma}\right)N_{a0}^-\int_0^h dz\left(\exp\left(\frac{q\varphi(z)}{k_BT}\right)-1\right)\end{pmatrix}. \qquad \text{(A.8)}$$

Under the condition of thick films, $h>>R_S$, Eq.(A.8) reduces to

$$u_3(V_0) \approx -\begin{pmatrix} \lambda\left(\tilde{\tilde{\Xi}}^V, -\tilde{\gamma}\right)\sqrt{\dfrac{2\varepsilon_{33}\varepsilon_0 k_B T}{q^2} N_{a0}^-}\left(\exp\left(-\dfrac{qV_0}{2k_B T}\right)-1\right) \\ +\lambda\left(\tilde{\beta}^a, \tilde{\gamma}\right)\sqrt{\dfrac{2\varepsilon_{33}\varepsilon_0 k_B T}{q^2} N_{a0}^-}\left(\exp\left(\dfrac{qV_0}{2k_B T}\right)-1\right) \end{pmatrix}. \quad \text{(A.9)}$$